\documentclass[a4paper,11pt]{article}
\usepackage{pos}
\usepackage{wrapfig}
\newcommand{\moonbeam}{\emph{MoonBEAM}}

\title{The Scientific Performance of the Moon Burst Energetics All-sky Monitor (MoonBEAM)}
\ShortTitle{MoonBEAM Science}

\author*[a]{C. Fletcher}
\author[b]{C. M. Hui}
\author[a]{A. Goldstein}
\author{the MoonBEAM Team}

\affiliation[a]{Science and Technology Institute, Universities Space Research Association, Huntsville, AL 35805, USA}

\affiliation[b]{NASA Marshall Space Flight Center, Huntsville, AL 35812, USA}

\emailAdd{cfletcher@usra.edu}

\abstract{\moonbeam\ is a SmallSat concept placed in cislunar orbit developed to study the progenitors and multimessenger/multiwavelength signals of transient relativistic jets and outflows and determine the conditions that lead to the launching of a transient relativistic jet. The advantage of \moonbeam\ is the instantaneous all-sky coverage due to its orbit, which maximizes the gamma-ray transient observations and provides upperlimits for non-detections. Earth blockage and detector downtime from the high particle activity in the South Atlantic Anomaly region prevent gamma-ray observatories in low Earth orbit from surveying the entire sky at a given time. In addition, the long baseline provided from a cislunar orbit allows \moonbeam\ to constrain the localization annulus when combined with a gamma-ray instrument in low Earth orbit utilizing the timing triangulation technique. We present the scientific performance of \moonbeam\ including the expected effective area, localization ability and duty cycle. \moonbeam\ provides many advantages to the gamma-ray and gravitational-wave follow up community by reducing the search region needed to identify the afterglow and kilanova emission. In addition, the all-sky coverage will provide insight into the conditions that lead to a successful relativistic jet, instead of a shock breakout event, or a completely failed jet in the case of core collapse supernovae.}

\ConferenceLogo{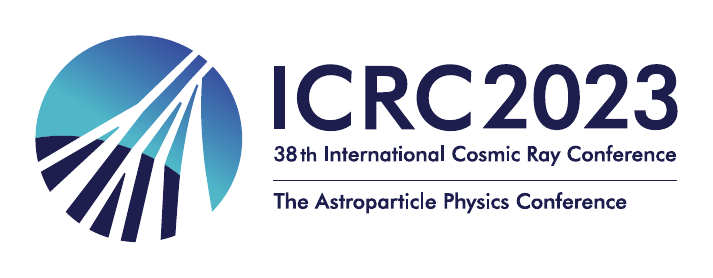}

\FullConference{%
38th International Cosmic Ray Conference (ICRC2023)\\
  26 July - 3 August, 2023\\
  Nagoya, Japan}

\begin{document}
\maketitle

\section{\moonbeam\ Overview}
The need for more gamma-ray instruments in orbit is critical for multimessenger and  multiwavelength astronomy. As the larger gamma-ray instruments currently in orbit are aging and with no replacement on the horizon, many missions concepts have turned to smaller form factors such as CubeSats and SmallSats.

Most gamma-ray instruments are placed on satellites in Low Earth Orbit (LEO) in order to detect gamma rays above the Earth's atmosphere. However, satellites in LEO go through an area of high particle activity called the South Atlantic Anomaly (SAA), which requires detectors to be turned off to avoid negative effects, reducing the instruments duty cycle. The Earth also blocks \~30\% of the field-of-view (FoV) for instruments in LEO, therefore if a gamma-ray transient were to occur behind the Earth the instrument would not be able to detect it. Furthermore, scattering off the Earth's atmosphere causes high background radiation and possible confusion of the gamma-ray transient's true location.

To solve the issues with instruments in LEO we present the  Moon Burst Energetics All-sky Monitor (\moonbeam) concept \citep{hui}. \moonbeam\ is a sensitive gamma-ray mission with instantaneous all-sky gamma-ray field of view capabilities. \moonbeam\ will be placed in a cis-lunar orbit allowing it to have a high-duty cycle (>98\%), little Earth blockage and relatively stable background. \moonbeam\ will be a nominal 2.5-year gamma-ray mission, with the possibility of extension, and the goal to observe gamma-ray transients from various progenitors (i.e.~binary compact mergers, core collapse supernovae, and magnetar giant flares) and enable very high energy gamma-ray and optical follow up campaigns.
\begin{figure*}[b]
\centering
\includegraphics[width = 0.90\textwidth]{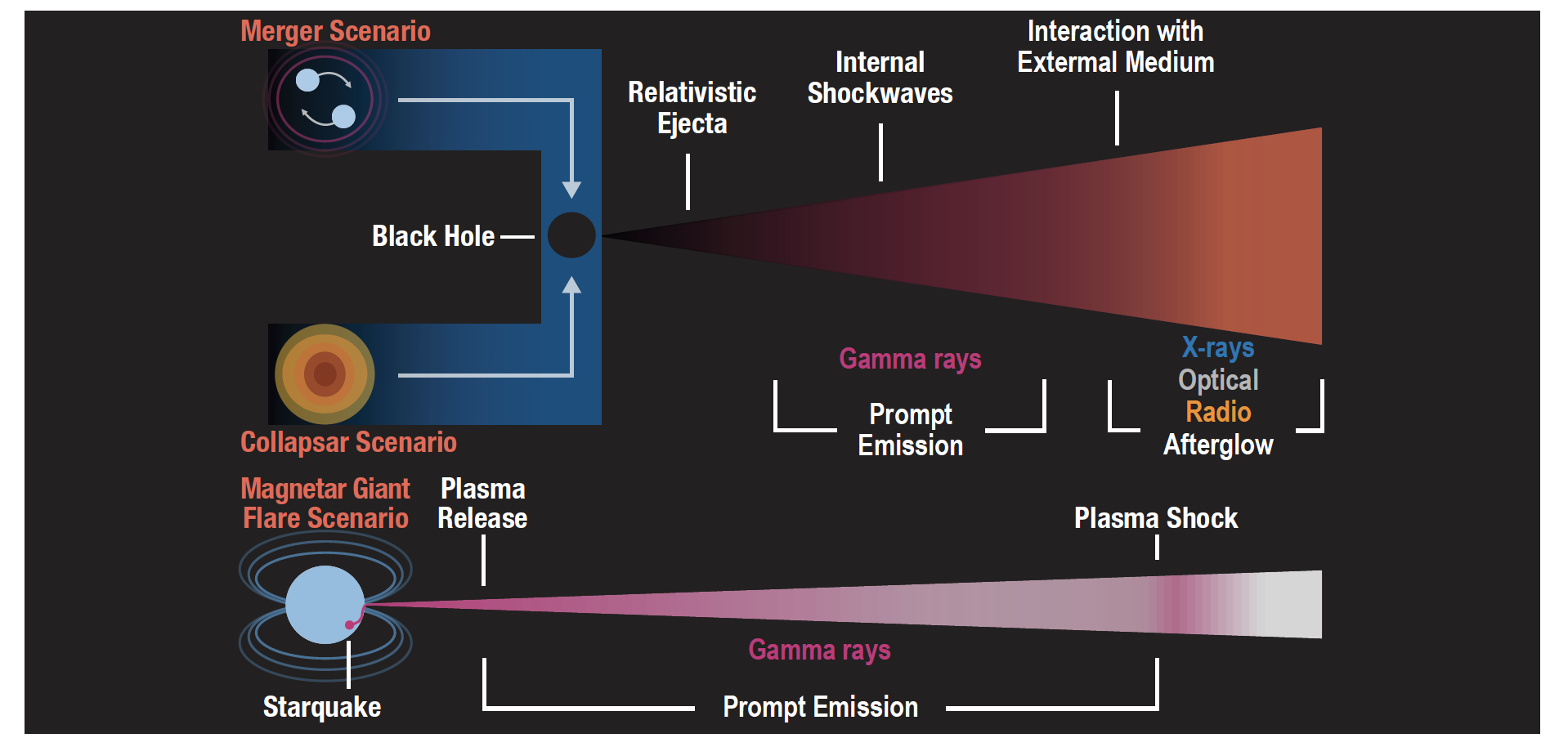}
\caption{Schematic of how a central engine that launches a relativistic outflow can be formed from three known progenitors, mergers of compact objects, collapsars, and magnetar giant flares. The prompt gamma rays probe the launch and emission mechanisms and provide the first notice that a relativistic transient has occurred. This initial notice allows for follow-up observations that cover the full EM spectrum providing complete picture of the process.}
\label{fig:progenitors}
\end{figure*}
The \moonbeam\ baseline mission will contain 6 scintillating detectors strategically placed on the spacecraft to provide an instantaneous all-sky field-of-view, unocculted by the Earth. Using a detector design of two materials, sodium iodide NaI(Tl) and cesium iodide CsI(NaI), \moonbeam\ will be sensitive to an energy range of 10–5,000 keV. The detector design utilizes a phoswhich configuration allowing for pulse shape discrimination (PSD) to distinguish in which scintillator a signal originated. The CsI(Na) component of the detector increases the energy range and effective area when compared to only NaI detectors. The detectors can also  be used to localize a gamma-ray event with PSD to ignore (veto) any signals from the CsI(Na), significantly improving localization \citep{wood}.

\section{Science Goals}
The science goals of \moonbeam\ are to explore the behavior of matter and energy under extreme conditions by observing relativistic astrophysical explosions. Progenitors of these transient bursts of emission have been confirmed to be the merger of two compact objects (neutron star or black-hole), a collapsar (type of core collapse supernova), or a giant flare generated by a starquake on a magnetar (a neutron star with
extremely powerful, large-scale magnetic field).
Relativistic transients produce
emissions across the electromagnetic
spectrum (EM) as well as multi-messenger
signals such as photons, gravitational
waves (GWs), neutrinos, and cosmic rays. Figure \ref{fig:progenitors} shows a schematic of these types of progenitors and how they produce not only prompt gamma-ray emission but long-lived multi-wavelength emission, and multi-messenger signals through relativistic jets and ejecta.
\begin{wrapfigure}{r}{0.5\textwidth}
\centering
\includegraphics[width = 0.48\textwidth]{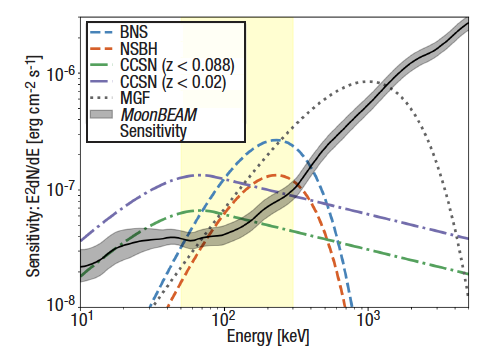}
\caption{The projected 4.5 $\sigma$ limiting. The sky-averaged sensitivity is the solid black line, and the gray band represents 99\% of the variability of the sensitivity on the sky. The dashed lines represent the candidate spectra and fluxes for each of the various sources using the baseline mission requirements in order to set sensitive upper limits. The yellow shaded region is the energy range (50–300 keV) over which the minimum flux sensitivity requirement is set.}
\label{fig:energy_sensitivity}
\vspace{-40pt}
\end{wrapfigure}

The science objectives of \moonbeam\ are
\begin{enumerate}
    \item Characterize the progenitors of gamma-ray bursts (GRBs) and their multi-messenger and multi-wavelength signals.
    \item Identify conditions necessary to launch a transient astrophysical jet.
    \item Determine the origins of the observed high-energy emission within the relativistic outflow.
\end{enumerate}
To achieve these objectives, \moonbeam\ aims to
determine the percentage of binary neutron star (BNS) mergers that produce jets and the resulting jet width, assuming a Gaussian-shaped jet. \moonbeam\ will also investigate the the percent of neutron star-black hole (NSBH) mergers that produce relativistic jets. Similarly, \moonbeam\ will examine the percentage of Core Collapse Supernovae (CCSNe) that produce jets and the percentage of those that produce a choked or failed jet. \moonbeam\ will also determine if magnetars can produce multiple magnetar giant flares (MGFs)\citep{roberts}. Furthermore, \moonbeam\ will enable optical follow-up of at least 300 GRBs and at least 10 very-high-energy (VHE) observations of GRBs within a redshift of 0.5 over its 2.5 year lifetime, based on GRB source rates from \cite{WP2010} and \cite{Howell2019}.

In the era of multimessenger astronomy,
simultaneous broadband observations are
vital for constructing a comprehensive picture
of relativistic transients and the outflows
they produce. Using joint multiwavelength and multimessenger observations to study the central engines that power these explosions is crucial to providing insights into
the composition of relativistic outflows, and
set strict constraints on the timescales for
jet formation and propagation. \moonbeam\
provides the essential continuous all-sky
gamma-ray observations that were identified as
a critical part of the Astro2020 Decadal Survey
need for the next decade in transient and multimessenger
astronomy \citep{astro2020}, by reporting any prompt emission of a relativistic transient, and by providing rapid
alerts to the astronomical community for contemporaneous and follow-up observations.

\section{Science Implementation}

The \moonbeam\ science goals encompass a large distance and luminosity range.
Observations of emissions from the
various progenitors span over eight orders of
magnitude in luminosity and a distance
ranging from nearby galaxies to
to the deaths of first-generation stars in the early universe. The ability to observe this dynamic range is critical to the science objectives.

\begin{wrapfigure}{r}{0.5\textwidth}
%centering
\includegraphics[width = 0.48\textwidth]{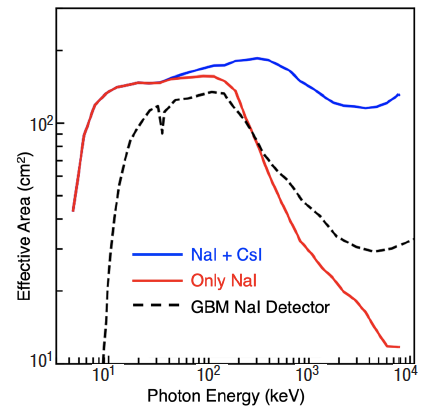}
\caption{The MoonBEAM effective area as a function of photon energy for a single phoswich detector using both NaI and CsI materials (blue ) is compared to using the phoswhich veto option of only the NaI material (red). A single Fermi-GBM NaI Detector is show in the dashed black line for reference.}
\label{fig:effarea}
\end{wrapfigure}
The limiting flux was determined for the gamma-ray transient progenitor types to provide context for \moonbeam's sensitivity. The different progenitors have a large range of spectral variation between, and also
within themselves, therefore the 50–300 keV energy range was appropriate to determine the limiting flux of these sources. For each of the scenarios in Figure \ref{fig:progenitors}, a candidate spectrum is used to calculate the limiting flux in the energy range 50–300 keV. The spectrum from GRB 170817A \citep{Goldstein2017}, the only confirmed GRB from a BNS merger, was used for the Merger Scenario. The spectrum from GRB 980425 \citep{Frontera2000}, confirmed to be
associated with a nearby CCSN, was
used for the CCSNe Scenario. The spectrum
of GRB 200415A \citep{Roberts2021}, a bright
extragalactic MGF from the Sculptor Galaxy,
was used as a candidate spectrum for the MGF Scenario.
Figure \ref{fig:energy_sensitivity} shows these spectra in comparison to the
projected performance of the \moonbeam\ mission. The figure illustrates that 50–300 keV is the optimal energy range over which to detect the various progenitors. \moonbeam\ will place sensitive limits on the flux of gamma-ray transient events in the case it does not detect the event but another instrument does.

The Medium-Energy Gamma-ray Astronomy
Library \citep{megalib}, which uses
Geometry and Data Tracking (Geant4) \citep{geant4}, was  used to
simulate gamma-ray interactions in a
model of the \moonbeam\ spacecraft. These
simulations, incorporating all the
major elements of the spacecraft, are used to determine various attributes of the science performance of the instrument, such as detector's effective area as a function of energy and direction, the sensitivity to GRBs, the intrinsic localization capability, and the
response matrices.
The simulations also provide the ability to examine the effectiveness of the phoswich veto mode and its impact on reducing the background rate and improving
\moonbeam's localization of gamma-ray
transients.

\begin{wrapfigure}{l}{0.5\textwidth}
\centering
\includegraphics[width = 0.48\textwidth]{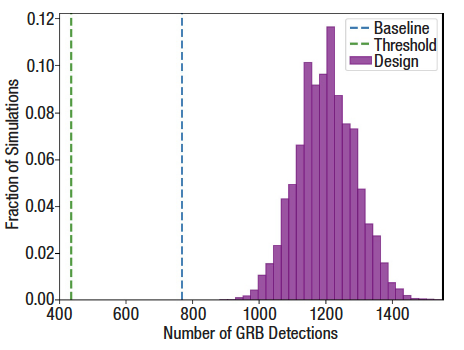}
\caption{ The predicted distribution of
detected GRBs for the \moonbeam\ mission timeframe. The baseline science requirement is the blue dashed line and threshold science requirement is green dashed line.}
\label{fig:grb_detections}
\end{wrapfigure}

The on-axis effective area of one detector
across the \moonbeam\ energy band is shown
in Figure \ref{fig:effarea} with a comparison to the phoswhich veto mode (Only NaI). The peak energy range of GRBs is between 50–300
keV \citep{vonKienlin2020}, therefore an instrument must have sensitivity over that range to successfully detect GRBs.  For the initial simulations, monoenergetic beams of gamma rays from 5 to 5,000 keV were used to determine the effective area across the sky. The average effective area at 300 keV is 645 cm$^{2}$ with minor fluctuations across the sky. The phoswich veto reduces the background incident on the rear of the detectors in this energy range and increases the angular dependence of the response for localization, justifying the phoswich design for localization.
\begin{wrapfigure}{r}{0.5\textwidth}
\centering
\includegraphics[width = 0.48
\textwidth]{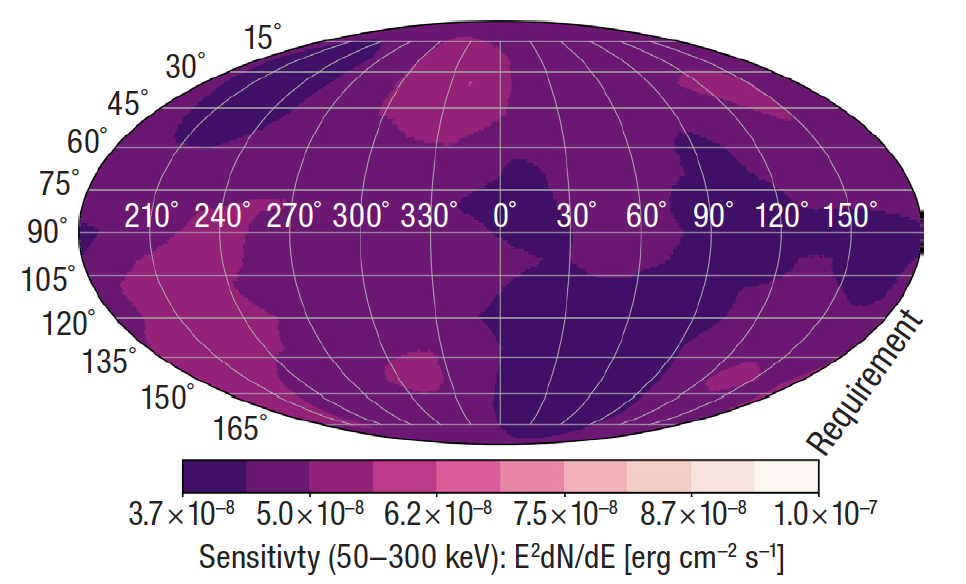}
\caption{MoonBEAM’s sensitivity
across the sky. The sensitivity requirement
is at the right most position in the scale.
MoonBEAM is at least 2 times more sensitive at
any point in the sky than this requirement.}
\label{fig:sky_sensitivity}
\end{wrapfigure}

\moonbeam\ is sensitive to a broad energy
range (10–5,000 keV) of photons and provides an energy resolution better than 12\% at 662 keV. Using 10 years of GRB detections reported by
the Fermi Gamma-Ray Burst Monitor (GBM) \citep{vonKienlin2020}, we predicted the detection distribution of the \moonbeam\ instrument by coupling its detection criteria and expected background rates. This GRB detection distribution is presented in Figure \ref{fig:grb_detections} for a time frame of 2.5 years of science operations. If \moonbeam\ does not detect a GRB observed by another instrument, such as mergers seen in gravitational
waves or supernovae detected in optical
wavelengths, it will provide unprecedented sensitive gamma-ray upper limits. Figure \ref{fig:sky_sensitivity} shows the limiting
flux sensitivity of \moonbeam\ across the entire
sky. \moonbeam\ provides a sensitivity improvement over current missions in LEO due to its  combined advantages of cislunar orbit and instrument design.

\begin{wrapfigure}{r}{0.5\textwidth}
\centering
\includegraphics[width = 0.48\textwidth]{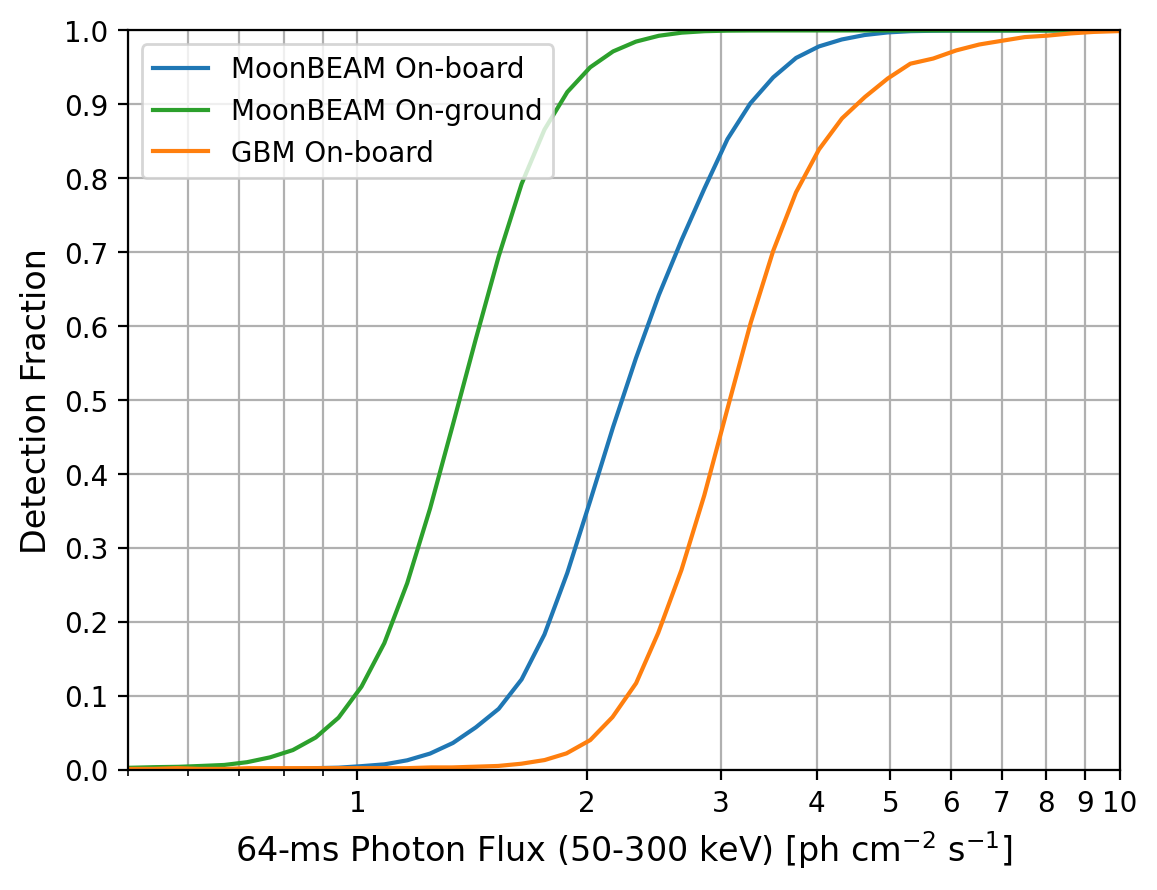}
\caption{ \moonbeam's threshold detection fraction of short GRBs as
a function of photon flux. The black filled circle is the expected performance. For comparison, the corresponding detection fraction curve is shown for Fermi GBM.}
\label{fig:detfraction}
\end{wrapfigure}

The predicted \moonbeam\ detection fraction of short GRBs detected as a function of 64-ms photon flux over 50–300 keV (the peak GRB energy range) was for both on-board and on-ground algorithms, shown in Figure \ref{fig:detfraction}. As expected, increasingly
bright, short GRBs are detected at higher
fractions. The figure compares the
detection fraction curve for Fermi GBM, calculated in the same manner except
with the observed Fermi GBM background
rate, to \moonbeam\ and clearly show that \moonbeam\ is more sensitive to
short GRBs than Fermi GBM. This is due to the
lower background and increased detector
size. For long GRBs the detection fraction was also calculated and is on par with Fermi GBM’s
performance. \moonbeam’s performance is overall better due to the stability of the background and longer possible integration times.

For gamma-ray transients, it is essential to provide a localization for every GRB detected to
enable multiwavelength/multimessenger follow-up. The localization informs
telescopes where on the sky to observe to find
a coincident counterpart \citep{Coughlin2019, Mong2021, ahumada2021} and provides information on whether two independent observations of a transient signal are associated. If a coincident counterpart is detected, the localization can be used to improve the localization even further.
In order to localize GRBs, \moonbeam\ uses a technique
\citep{Connaughton2015, Goldstein2020}
employed by Fermi GBM that generates count
rate “templates” on a grid on the sky, assuming
a particular GRB spectrum, for each of the
detectors. These templates encode
the expected relative count rates between each
detector as a function of arrival direction and the observed count rates are compared to the
templates at each point on the sky grid to compute a probability map.

We examined the \moonbeam\ localization capability by simulating random GRBs sampled  from the 10-year Fermi GBM spectral catalog \citep{Poolakkil2021}. Each randomly selected GRB is
then assumed to arrive from a random direction,
for which the \moonbeam\ responses are
generated to convert from photons to observed
counts in the detectors. Following
the same procedure, the localization capability
of the phoswich mode was evaluated by using
detector responses made from accepting photons
that interact only in the NaI(Tl) (i.e., the CsI(Na)
is used as a veto). Fifty percent
of the ensemble of GRBs are located with a 1$\sigma$
statistical error of 4.5$^\circ$. A conservative systematic uncertainty
of 3$^\circ$ is estimated from Fermi GBM \citep{Connaughton2015, Goldstein2020} and increases
the total estimated localization uncertainty to
5.4$^\circ$, which is comparable to Fermi GBM.

\begin{wrapfigure}{r}{0.5\textwidth}
\centering
\includegraphics[width = 0.48\textwidth]{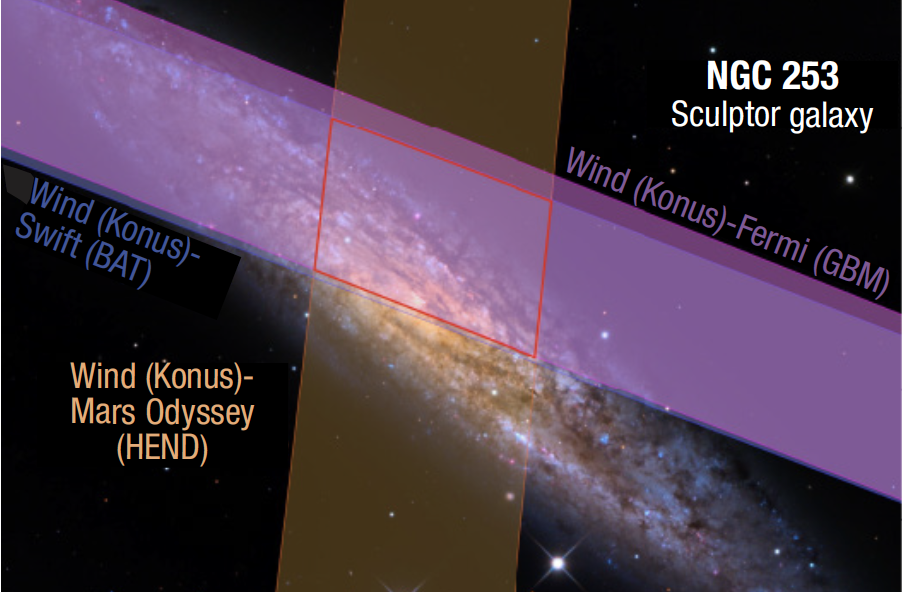}
\caption{IPN localization of GRB
200415A, determined to be an MGF. The IPN used three gamma-ray instruments and found that the location
of the MGF was the Sculptor Galaxy.}
\label{fig:ipn}

\end{wrapfigure}

\moonbeam\ will join the Interplanetary Gamma-Ray Burst Timing Network\footnote{https://heasarc.gsfc.nasa.gov/docs/heasarc/missions/ipn.html} (IPN) which combines the data from instruments both in and outside of LEO to achieve an improved localization though a timing triangulation method for any high-energy transients \citep{Hurley2013}. The current IPN consists of eleven missions, three of
which are at interplanetary distances (Konus
GRB instrument on Wind; Mars Odyssey;
and BepiColombo). \moonbeam\ would provide the IPN with the only dedicated GRB mission at
> 150,000 km from the Earth, launched within
within the past 20 years, and the only one outside of LEO with the capability of on-board transient localization. Figure \ref{fig:ipn} shows an example of how the IPN was used to determine the host galaxy for GRB 200415A, a MGF masquerading as a short GRB.

\section{Summary}
\moonbeam\ provides essential gamma-ray observations of relativistic astrophysical transients with the following capabilities: instantaneous all-sky field of view from a lunar resonant orbit, >98\% duty cycle, relatively stable background, sensitivity to prompt GRB emission energy range and broad coverage for spectroscopy. \moonbeam\ also provides independent localization as well as a longer baseline for additional localization improvement with other gamma-ray missions, and rapid alerts to the astronomical community for contemporaneous and follow-up observations.
\moonbeam's projected observations are 4,000+ gamma-ray transients with either detections or sensitive upper limits, enabling > 400 follow-up observations, > 1000 GRB detections, and all-sky limiting flux is 2x more sensitive than required to determine jet production rate in different progenitors. In the era of multimessenger astronomy instantaneous all-sky gamma-ray instruments are imperative. The full \moonbeam\ Team author list can be found in \citep{hui}.

\bibliographystyle{aasjournal}
\bibliography{bib}

% % Full authors list (ONLY FOR COLLABORATIONS)
% \clearpage
% \section*{Full Authors List: \Coll\ Collaboration}

% \noindent \textbf{Note comment afterwards:} Collaborations have the possibility to provide an authors list in xml format which will be used while generating the DOI entries making the full authors list searchable in databases like Inspire HEP. \\

% \scriptsize
% \noindent
% first.author$^1$,
% second.author$^2$,
% third.author$^3$ % .... more names
% and
% last.author$^{n}$ \\

% \noindent
% $^1$first.affiliation.
% $^2$second.affiliation. % .... more affiliation
% $^{m}$last.affiliation.

\end{document}